\documentclass[prb,twocolumn,showpacs]{revtex4}

\usepackage{graphicx}
\usepackage{dcolumn}
\usepackage{amsmath}

\begin{document}
\def\btau{\mbox{\boldmath$\tau$}}
\def\bmu{\mbox{\boldmath$\mu$}}
\def\bsigma{\mbox{\boldmath$\sigma$}}

\title{Damping of spin waves and singularity of the longitudinal modes in the dipolar critical regime of the Heisenberg-ferromagnet EuS}
\author{P.~B\"oni}
\affiliation{Physik-Department E21,
        Technische Universit\"at M\"unchen,
        D-85748 Garching, Germany  \\and
        Laboratory for Neutron Scattering ETH \& PSI,
        CH-5232 Villigen PSI, Switzerland}
\author{ B.~Roessli }
\affiliation{Institut Laue Langevin ,
        F-38042 Grenoble, France \\ and
        Laboratory for Neutron Scattering ETH \& PSI,
        CH-5232 Villigen PSI, Switzerland}
\author{D.~G\"orlitz}
\author{J.~K\"otzler}
\email{koetzler@physnet.uni-hamburg.de}

\affiliation{Institut f\"ur Angewandte Physik und 
        Zentrum f\"ur Mikrostrukturforschung  \\
        D-20355 Hamburg, Germany}

\date{October 17, 2001}
\begin{abstract}
By inelastic scattering of polarized neutrons near the (200)-Bragg reflection, the susceptibilities and linewidths of the spin waves and the longitudinal spin fluctuations, $\delta {\bf S}_{sw}({\bf q})$ and $\delta {\bf S}_{z}({\bf q}) \parallel {\bf M}_s$, respectively, were determined separately. By aligning the momentum transfers ${\bf q}$  perpendicular to both $\delta {\bf S}_{sw}$ and the spontaneous magnetization ${\bf M}_s$, we explored the statics and dynamics of these modes with transverse polarizations with respect to {\bf q}. In the dipolar critical regime, where the inverse correlation length $\kappa_z(T)$ and q are smaller than the dipolar wavenumber $q_d$, we observe: (i) the static susceptibility of $\delta {\bf S}_{sw}^T({\bf q})$ displays the Goldstone divergence while for $\delta {\bf S}_z^T({\bf q})$ the Ornstein-Zernicke shape fits the data with a possible indication of a thermal (mass-) renormalization at the smallest q-values, i.e. we find indications for the predicted 1/q divergence of the longitudinal susceptibility; (ii) the  spin wave dispersion as predicted by the Holstein-Primakoff theory revealing $q_d=0.23(1)~\AA^{-1}$ in good agreement with previous work in the paramagnetic and ferromagnetic regime of EuS;  (iii) within experimental error, the (Lorentzian) linewidths of both modes turn out to be identical with respect to the $q^2$-variation, the temperature independence and the absolute magnitude. Due to the linear dispersion of the spin waves they remain underdamped for $q <q_d$. These central results differ significantly from the well known exchange dominated critical dynamics, but are quantitatively  explained in terms of dynamical scaling and existing data for $T \ge T_C$. The available mode-mode coupling theory, which takes the dipolar interactions fully into account, describes the gross features of the linewidths but not all details of the T- and q-dependencies.
\end{abstract}
\maketitle
\section{Introduction}
Neutron scattering has been demonstrated as an extremely useful probe of the spin fluctuations $\delta {\bf S}(q)$ near the Curie temperature of Heisenberg ferromagnets \cite{collins,minciewicz,dietrich,bohn}. Under the assumption that the {\em isotropic exchange interaction} dominates the ordering process, the early results could be well interpreted in terms of the dynamical scaling hypothesis \cite{halperin}. In its simplest form this hypothesis states that as for the static spin correlations also the temperature and ${\bf q}$-dependencies of their characteristic frequencies are described by homogeneous scaling functions that depend only on a single variable,  $q\xi(T)$, where $\xi$  denotes the correlation length of the order parameter fluctuations. 

\begin{figure}
\includegraphics*[width=7.5cm]{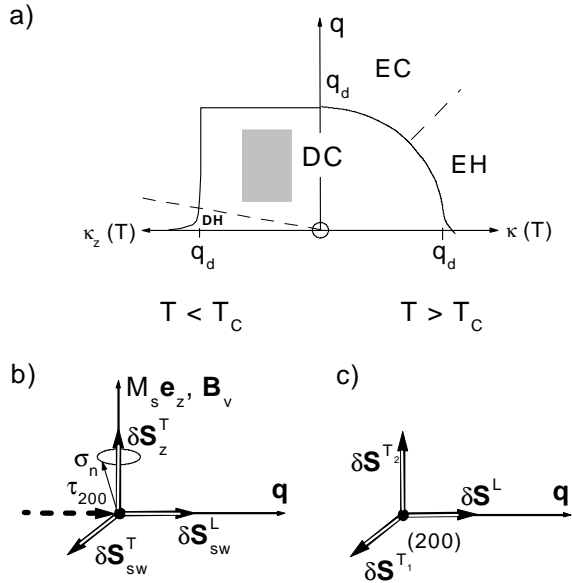}
\caption{\protect { a) Dipolar regime for static critical behavior below $T_C$ explored in the present work (shaded area), with $q_d=0.23 ~{\rm \AA}^{-1}$ for EuS. The dotted lines define the exchange (EH) and dipolar (DH) hydrodynamic critical regimes above and below $T_C$, respectively. b) Spin fluctuation modes defined by the geometry of our polarized neutron experiment below $T_C$ of EuS, where the momentum transfer ${\bf q}={\bf Q}-{\bf \btau}_{200}$ is kept perpendicular to the spontaneous magnetization. Spin wave $(\delta{\bf S}_{sw})$ and longitudinal $(\delta{\bf S}_z)$ fluctuations are detected separately by the spin-flip and non spin-flip intensities for neutrons with incident polarization ${\bsigma}_n$ and wave vector ${\bf Q}$. c) Definition of the magnetic modes with respect to the reduced momentum transfer {\bf q}
} }
\label{geometry}
\end{figure}

Signatures of the inevitable, {\em anisotropic dipole-dipole interaction} on the fluctuations of Heisenberg ferromagnets have been first detected by measurements of the relaxation rate  $\Gamma(0)$ of the homogeneous  $\delta S(0)$    mode in the {\em paramagnetic} phase of $CdCr_2Se_4$ \,\, \cite{philipsb}  and subsequently also of Fe by neutron spin-echo measurements at small q \,\, \cite{mezei83}. Based on these signatures and also on first theoretical approaches, which treated the dipolar interaction as a perturbation of the isotropic fluctuations \cite{huber,finger}, it has been conjectured \cite{Koe83} that the dipolar forces should gain importance somewhere in the socalled dipolar critical  (DC) regime, where  $\xi^{-1}\equiv \kappa$ and q are small compared to the dipolar wavenumber $q_d$ (see Fig.\ref{geometry}a). For a given ferromagnet, this quantity measures the strength of the dipolar interaction relative to the exchange interactions. It has been introduced in renormalization group (RG) calculations of the static correlation functions \cite{fisher73},  recognizing that the dipolar anisotropy breaks the rotational invariance of the fluctuations   $\delta{\bf S}({\bf q})$ with respect to ${\bf q}$. 
%Here, {\bf q} denotes the momentum transfer with respect to the nearest Bragg peak (see Fig. \ref{geometry}b). 
The demagnetizing effect of the dipolar interaction on the longitudinal modes  $\delta {\bf S}^{L} \parallel {\bf q}$, prevents them from becoming critical, while the remaining two transverse modes (see Fig.\ref{geometry}c) are driving the ferromagnetic transition. At first these dipolar anisotropic fluctuations have been realized  by elastic scattering of polarized neutrons above $T_C$ of the Heisenberg ferromagnets EuO and EuS \cite{koetzler86a}, where for the latter it was also possible to measure directly the characteristic wavenumber, $q_d = 0.22(5)\AA^{-1}$. 

On the theoretical side, the implications of the dipolar critical fluctuations on their dynamics  have been fully taken into account only by the mode-mode coupling (MMC), approach \cite{frey88,freylongpaper,schinz98b,schinz98c}. {\em Above $T_C$}, rather convincing agreement was obtained \cite{frey88,freylongpaper} for the critical slowing down of $\Gamma^\alpha({\bf q})$ observed for $q \to 0$ and $T \to T_C$ on the transverse fluctuations of Fe  \cite{mezei83} and EuS \cite{boni91} as well as for the longitudinal ones ($\alpha=L$) of EuS \cite{boni91}.  Additional strong support for the MMC-results came from quantitative analyses of the relaxation rates $\Gamma^T({\bf q},T)$ in the two limiting cases $q=0,T\agt T_C$ and $T=T_C, q\agt 0$ for the archetype Heisenberg ferromagnets \cite{koetzler88}.  As one of the striking  results we mention the crossover from $\Gamma^T(q>q_d,T_C) \sim q^{5/2}$ in the exchange dominated regime to $\Gamma^T( q \ll q_d,T_C) \sim q^2$, deep in the dipolar one \cite{mezei83,frey88,freylongpaper}. The latter behavior corresponds to the conventional (van Hove type) slowing down, characterized by a non-critical Onsager coefficient of the spin dynamics, $L^T({\bf q},T) \equiv \Gamma^T({\bf q})\cdot \chi^T({\bf q}) $\,\, \cite{frey88,freylongpaper, koetzler88}, where $\chi^T({\bf q})$ is the static susceptibility of the transverse fluctuations. It was shown that this central quantity depends in a universal manner only on $T_C$ and $q_d$, to be discussed in Sect.\ref{discussion}.

{\em Below $T_C$}, the situation becomes more complicated because the symmetry of the fluctuations is further reduced by the appearance of the order parameter, i.e. the spontaneous magnetization ${\bf M}_s$. As illustrated in Fig.\ref{geometry}b, it is common sense to distinguish there between the spin waves $\delta {\bf S}_{sw}({\bf q}) \perp M_s{\bf e}_z$ and the longitudinal modes $\delta {\bf S}_{z}({\bf q}) \parallel M_s{\bf e}_z$. A systematic classification of critical behaviors in the $q-\kappa(T)$-plane, which considers the Heisenberg exchange and the dipolar interaction on an equal footing, has recently been performed by Schinz and Schwabl \cite{schinz98a}. According to their results, we have depicted the dipolar critical (DC) regime in Fig.\ref{geometry}a, which apart from a dipolar hydodynamic (DH) regime at rather small q is bound from above by the dipolar wavenumber $q_d$   in q-direction as well as in the $\kappa_z(T)$-direction.

Since among the archetypical Heisenberg ferromagnets  $q_d$ is largest for EuS \cite{Koe83,koetzler88}, this material is preferred for experimental studies of dipolar effects \cite{koetzler86a,boni91,KKW76,boni95,sacavem}. Here we report results of a first systematic study of the magnetization dynamics in the DC-region, which covers the shaded area in Fig.\ref{geometry}a. Our principal goal is to determine the susceptibilities and linewidths of the spin waves and of the longitudinal fluctuations and to examine their q- and temperature variations. The experimental access by means of inelastic scattering of polarized neutrons around a finite Bragg peak is described in Section \ref{experiment}.
As illustrated by Fig.\ref{geometry}b we choose a configuration where the momentum transfer {\bf q} occurs in directions perpendicular to the order parameter
${\bf M}_s$, which allows to define and determine the transverse polarizations of the spin wave and of the longitudinal modes with respect to $\bf q$.
This will turn out to be essential for the discussion by means of dynamical scaling, because the appearance of the order parameter does not lift
the symmetry with respect to $\bf q$ so that both modes retain their transverse polarization from above $T_C$.
In Section \ref{results}, we give some examples for constant {\bf Q}-scans and the analysis of the inelastic cross section. We evaluate there the relevant observables, i.e. static susceptibilities, the spin wave frequencies and the linewidths. Their detailed temperature and q-variations are presented in Sect. \ref{analysis}. In particular, we determine there the dipolar wavenumber $q_d$ from spin wave energies and the correlation length $\xi(T)=\kappa_z^{-1}(T)$ of the longitudinal fluctuations from their static susceptibilities. Backed with these findings, we discuss the central results. i.e. the damping of the spin waves and of the longitudinal fluctuations, in Sect.\ref{discussion}. Here the objectives are twofold. First, basing on the existing data for the paramagnetic side of EuS we want to examine whether and how the scaling hypothesis, which is rather special for the dipolar interaction \cite{halperin}, works, and second we will compare our results to the rather detailed predictions of the recent numerical solutions of MMC equations \cite{schinz98c}. To produce explicit values for the linewidth, this MMC approach had to introduce several assumptions which, of course, need to be checked by the experiment. The paper closes with a brief summary and outlook.

\begin{figure}
\includegraphics*[width=7.5cm]{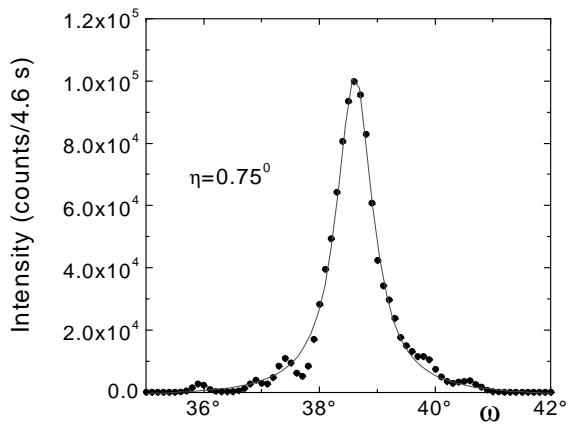}
\caption{\protect { Rocking curve of the $\btau_{200}$ Bragg peak of the isotopically
    enriched sample $^{153}$EuS that is
    composed of more than 100 individual single crystals. The mosaic of the
    sample is $\eta = 0.75^\circ \pm 0.02^\circ$.      }}
\label{mosaic}
\end{figure}

\section{Experiment}
\label{experiment}

The neutron scattering experiments were performed using the triple
axis spectrometer IN14 at the ILL in Grenoble, with polarization
analysis. The isotopically enriched sample $^{153}$EuS was
composed of roughly 100 single crystals, aligned such that the
overall mosaic was $\eta \simeq 0.75^\circ$ (see Fig.~\ref{mosaic}).
In EuS the magnetic $Eu^{2+}$ ions ($S=7/2, g_L=2$) are arranged on a face centered cubic
lattice, $a_0= 5.973 \AA $.
The sample was mounted inside a superconducting magnet providing
vertical fields $B_v$ up to 4.5~T. All measurements were conducted near
the (200) Bragg peak with ${\bf q}$ along the [h00] direction (see Fig.\ref{geometry}b)
using neutrons with fixed incident
energies $E_i = 3.0$ and 3.8 meV and collimations 37'-37'-40'
along the beam direction from the monochromator to the detector. Due to the mosaic structure of the sample shown in Fig.\ref{mosaic} the measurements in the other {\bf q}-direction suffered from a high background from the elastic Bragg intensity.

The neutron cross section for neutrons scattered from an
isotropic magnetic material is given  by
\cite{lovesey87}
\begin{eqnarray}
\label{crose}
{d^2\sigma\over d\Omega dE'} = 
{k_f \over k_i}
     \biggl( {\gamma r_0\over2}\biggr)^2 N  F({\bf Q})^2 \exp(-2W({\bf Q})) \times
\nonumber
 \\
     \sum_{\alpha\beta} (1 - \hat Q_{\alpha\beta}^2) S_{\alpha\beta}(\bf q,\omega).
\end{eqnarray}
${\bf Q = k}_i - {\bf k}_f$ and $\hbar \omega = E_i - E_f$ are the total
momentum and energy transfers from the neutron to the sample,
respectively, where $E_{i,f} = (\hbar k_{i,f}^2/(2m_n)$ is the
neutron energy. $\alpha$ and $\beta$ designate the cartesian
components $x,y,z$, $(\gamma r_0/2)^2 = 72.65 \cdot 10^{-3}$
barn/$\mu_B^2$, $F$ is the magnetic form factor for the Eu$^{2+}$
ion, the exponential term is the Debye-Waller factor, $ \hat {\bf Q}~=~{\bf Q}/|{\bf Q}|$,
 and $N$ is the number of the magnetic ions in the
sample. The reduced momentum transfer is defined by ${\bf q} = {\bf Q} -
\btau_{200}$, where $\btau_{200}$ is the position of the nearest Bragg peak, 
$\btau_{200}=(200)2\pi / a_0$ in our case, (Fig.~\ref{geometry}b).

The $\alpha\beta$-component of the scattering function
$S_{\alpha\beta}(\bf q,\omega)$ is related to the imaginary part
of the susceptibility by the fluctuation-dissipation theorem
\[S_{\alpha\beta}({\bf q},\omega) = \langle n+1 \rangle {1\over \pi} \Im\chi_{\alpha\beta}({\bf q},\omega),\]
where the thermal population factor is given by $\langle n \rangle =[\exp(\hbar \omega / k_BT)-1]^{-1}$.
Our experiments have been conducted close to the Curie
temperature $T_C=16.25~K$ and at energy transfers $\hbar \omega \ll k_BT$,
so that $S_{\alpha\beta}$ becomes directly proportional to
$T\Im\chi_{\alpha\beta}$
\begin{equation}
\label{fdtsimple}
 S_{\alpha\beta}({\bf q},\omega) = {k_BT \over \hbar\omega} {1\over \pi} \Im\chi_{\alpha\beta}({\bf q},\omega).
\end{equation}
Thus, the neutron scattering cross section reflects directly the
$\bf q$- and $\omega$-dependence of the susceptibility components of the sample.

According to Eq.(\ref{crose}), the neutrons couple only to spin 
fluctuations $\delta{\bf S}$ that are perpendicular
to the scattering vector ${\bf Q} = \btau_{200} +{\bf q}$. Therefore, the cross section contains
contributions from the longitudinal fluctuations $\delta {\bf S}_z^T$ and from the
spin wave scattering that corresponds to excitations with $\delta
{\bf S}_{sw}^T$ perpendicular to $\bf M$. Both modes have a
transverse polarization with respect to ${\bf q}$, and in the following we omit the superscript T.
The fluctuations parallel to $\bf M$ can be separated from the spin wave
modes by analyzing the polarization ${\bsigma}_f = \pm {\bsigma}_i$
of the scattered neutrons with respect to the incident polarization, ${\bsigma}_i \parallel {\bf B}_v$.
As can be inferred from Fig.\ref{geometry}b, the parallel fluctuations give rise to non-spin flip
scattering and the spin wave modes give rise to spin flip
scattering. Therefore, the susceptibilities $\chi_z({\bf q},\omega)$ and $\chi_{sw}({\bf q},\omega)$
of both magnetic modes, can be determined unambiguously in a vertical field ${\bf B}_v$. 
 In order to minimize the influence
of the magnetic field on the spin fluctuations as much as possible,
we adjusted ${\bf B}_v$ so that the internal magnetic field ${\bf B}={\bf B}_v - N_z{\bf M}(T)$,
with $N_z=0.05$, was small, however large enough to remove the domain walls. The flipping ratio was 
between $2.1 < R < 3.5$. 

In a first step, all the measured data $I_{\rm obs}^{\mu\nu}$ was corrected for the
finite flipping ratio $R$ according to:
\[ I_{z} = {R\over R-1} I^{++}_{\rm obs} - {1\over R-1} I^{+-}_{\rm obs},\]\\
\[ I_{sw} = {R\over R-1} I^{+-}_{\rm obs} - {1\over R-1} I^{++}_{\rm
 obs},\]
where $I^{\mu\nu}$ designates the scattered intensity from
the polarization $\mu$ to the polarization $\nu$ of the
incident and scattered neutrons, respectively.
In a second step, a background being determined in the paramagnetic
phase at $T = 80$ K $\gg T_C$ and in the ordered phase at $T =1.78$ K, 
was subtracted from the data. The energy-independent
contribution was 1 count/8.3 min for all the measurements in the
range $0.06 < \zeta < 0.18$ \,\,\cite{background,boni87}, where the reduced
momentum transfer  is measured in reciprocal lattice units $\zeta=q/(2\pi/a_0)$.
The peak intensity of the elastic background was 4 counts/8.3 min
and 8 counts/min for the spin-flip and non-spin-flip scattering at
the position $\btau + {\bf q}$, respectively. 

\begin{figure}
\includegraphics*[width=7.5cm]{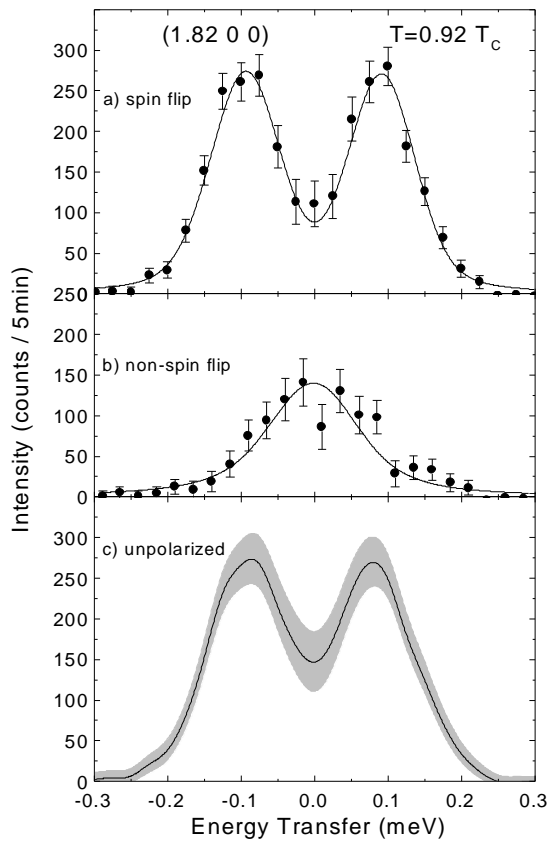}
\caption{\protect { Constant-Q scans for a) spin flip and b) non-spin flip neutrons scattered from spin wave and longitudinal fluctuations, respectively, at $T=0.92T_C=15.0~K$. c) Spectrum for unpolarized neutrons, calculated from a) and b), showing the depression of the longitudinal contribution.      }}
\label{Qscan18}
\end{figure}

\section{Constant-Q Spectra}
\label{results}

The inelastic magnetic scattering has been determined at several
temperatures and momentum transfers by performing constant-$\bf Q$ scans.
Fig.~\ref{Qscan18} shows the cross sections
at $\zeta = 0.18$ as measured along the [100] direction at the
(2 0 0) Bragg reflection at $T = 15$ K in a field $ B_v =100$ mT. 
The spin-flip data clearly reveals spin waves, while the non-spin flip data is quasielastic and
has a width (half-width at half maximum) that is roughly a factor of two smaller than the energy of the spin waves. Similar data was collected for
many different temperatures $1.78 \le T \le 80$ K and in appropriate
fields 30 mT $\le  B_v \le$ 500 mT, as described in Section \ref{experiment}.

Scattering of polarized neutrons is a very efficient way to properly
separate the longitudinal from the transverse excitations in an
Heisenberg ferromagnet. This is demonstrated in Fig.~\ref{Qscan18}c
where we show for comparison the spectrum calculated from
the (flip efficiency) corrected intensities of Fig.~\ref{Qscan18}a,b:
\[   I_{unpol}(\omega) = {2\over 3} I_{sw}(\omega) + {1\over 3} I_z(\omega)\]
as it would be measured by means of unpolarized neutrons. The
comparison shows clearly that the longitudinal scattering cannot
be reliably extracted by means of unpolarized neutrons because the
spectral widths of $I_{sw}$ and $I_z$ are similar. Moreover,  reliable 
positions and widths of the spin waves can
only be obtained if polarized beam data is used.

\begin{figure}
\includegraphics*[width=7.5cm]{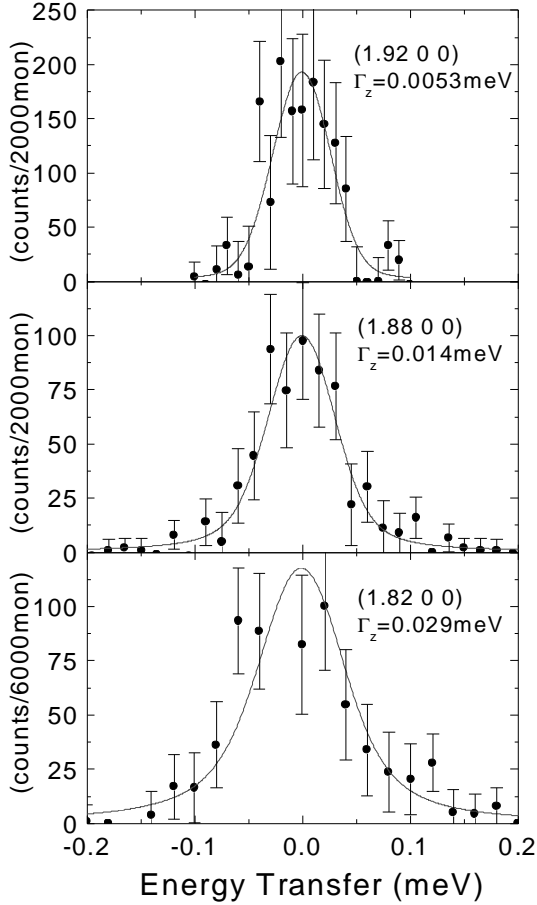}
\caption{\protect { Spectra of the longitudinal fluctuations recorded at $T=0.96 T_c$ and different momentum transfers ${\bf q}$. }}
\label{QdepPFT}
\end{figure}

In order to analyze the data we have employed the scattering function,
Eq.(\ref{crose}), which via Eq.(\ref{fdtsimple}) is directly related to
the dynamical susceptibilities $\chi_z$ and $\chi_{sw}$. 
To allow for a comparison between our results and the existing theory \cite{frey88},
we assume Lorentzian spectral weight functions for the dynamic
susceptibilities of the modes $\mu=z,sw$ and obtain

\begin{eqnarray}
\label{fitfun}
 \Im\bigl( \chi_\mu({\bf q},\omega)\bigr) = \chi_\mu({\bf q}) {1\over 2\pi}
 \biggl( {\hbar \omega \Gamma_\mu \over (\omega-\omega_\mu({\bf q}))^2 + (\Gamma_\mu)^2}
\nonumber
\\
 +{\hbar \omega \Gamma_\mu \over (\omega+\omega_\mu({\bf q}))^2 + (\Gamma_\mu)^2}
 \biggr),
\end{eqnarray}
where $\hbar \omega_{sw}\equiv E_{sw}({\bf q})$ is the spin wave energy and
$\Gamma_\mu({\bf q})$ are the linewidths. As the scattering from the longitudinal
fluctuations proves to be quasielastic, 
we set $\omega_{z}({\bf q}) = 0$.
The data was fitted by convoluting the scattering functions
$S_\mu({\bf q},\omega)$ with the four-dimensional
resolution function of the spectrometer to give $I_\mu({\bf q},\omega)$. The three parameters
$E_{sw}, \Gamma_{\mu}, \chi_{\mu}({\bf q})$ and a common
normalization parameter were varied for each constant-{\bf Q} scan
such that $\chi^2$ was minimized.
The solid lines in Fig.~\ref{Qscan18}a,b are fits to the data using
Eq.~(\ref{fitfun}). They describe the data well. 
Figure~\ref{QdepPFT} shows that the width of the longitudinal
fluctuations $\Gamma_z$ increases with increasing $q$ 
in qualitative agreement with existing theories \cite{halperin,huber,freylongpaper}. A more detailed comparison will be
performed in the discussion. 

\section{Analysis of the Spectra}
\label{analysis}

From the fits of the spectra using the double Lorentzian
scattering function, Eq.~(\ref{fitfun}), as described in the previous
paragraph, we extracted the static susceptibilities $\chi_\mu({\bf q},\omega=0,T) \equiv \chi_\mu({\bf q},T)$, the energies $E_{sw}({\bf q})$ and linewidths
$\hbar \Gamma_{sw}({\bf q})$ of the spin waves, and the linewidth
$\hbar \Gamma_{z}({\bf q})$ of the quasielastic scattering.
These quantities will be discussed in the following.

\subsection{Static Critical behavior}
\label{statsus}

In order to compare the experimentally determined static
susceptibilities with the theory we refer to the
expressions from the theoretical work by Schinz and Schwabl \cite{schinz98a}
who have presented $\chi_{\mu}({\bf q},T)$ for all q-values and temperatures below $T_C$.
Especially, one finds for the susceptibilities $\chi_{sw}$ and $\chi_z$, deep in the dipolar regime as covered
by our experiment (Fig.\ref{geometry}a):
\begin{equation}
\chi_{sw}({\bf q},T) = \frac{q_d^2}{q^2},
\label{chisw}
\end{equation}

\begin{equation}
\chi_{z}({\bf q},T) = \frac{q_d^2}{q^2+\kappa_-^2({\bf q},T)}.
\label{chiz}
\end{equation}

\begin{figure}
\includegraphics*[width=7.4cm]{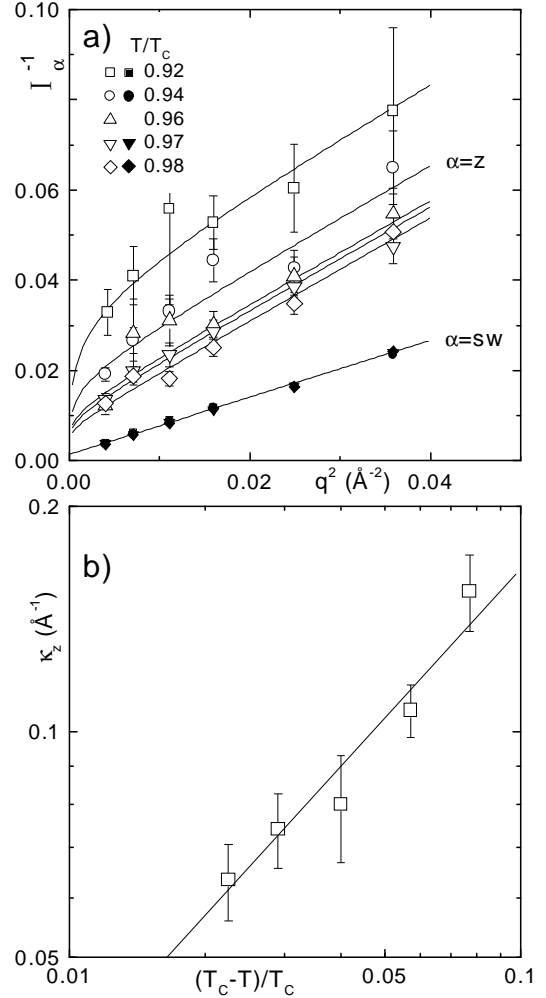}
\caption{\protect {a) Inverse of the integrated intensities of the spin wave and longitudinal spin fluctuations versus $q^2$, fitted to the  inverse static susceptibilities, $\chi_{\rm sw}^{-1} \sim (q^2+\kappa^2_g)$ and $\chi_z^{-1}(q\to 0) \sim (q^2+\kappa_-^2(T,q))$, Eq.(\ref{chiz}) b) Temperature dependence of the inverse correlation length of the longitudinal magnetization fluctuations $\kappa_z(T) = \xi^{-1}(T)$, defined by Eq.(\ref{kappam}a).}}
\label{invint}
\end{figure}

The $q^{-2}$-divergence is characteristic of gapless spin waves with transverse polarization 
$\delta{\bf S}_{sw} \perp {\bf q}$, also referred to as Goldstone modes, while the fluctuations
parallel to $\bf M$ acquire a (thermal) mass, $\kappa_-^2(T)$. At small momentum transfers
$q \ll \kappa_-$, this term is renormalized by the spin wave fluctuations and becomes q-dependent
\addtocounter{equation} {-1}
\begin{subequations}
\begin{equation}
\kappa_-^2({\bf q},T)= \frac{29}{18} \cdot \frac{\kappa_z^2(T)}{1+a \kappa_z(T)/q}, 
\end{equation}
\label{kappam}
\end{subequations}
where $\kappa_z=\xi^{-1}$ denotes the inverse correlation length below $T_C$ and $a \simeq 2/9$ \, \cite{schinz98a,mazenko}.
One consequence of this effect has been emphasized for the homogeneous susceptibility
already in the original spin wave work by Holstein and Primakoff \cite{HP40}.
They predicted the singularity for vanishing magnetic field, $\chi_z(q=0,T,B \to 0) \sim B^{1/2}$, 
which in fact has been confirmed in experiments on EuS and EuO \cite{koetzler94}. More recently, the crossover
from the Ornstein-Zernicke type behavior of the zero-field susceptibility to a $\chi_z(q \ll \kappa_z) \sim q^{-1}$
singularity has been obtained by the RG theory \cite{mazenko}, but a clear experimental evidence is yet
lacking.

\begin{figure}
\includegraphics[width=7.5cm]{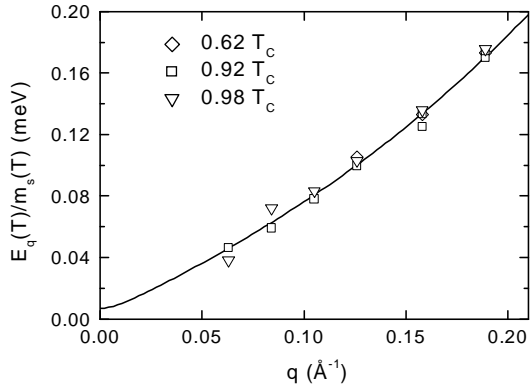}
\caption{\protect { Dispersion of the spin waves with transverse polarization $\delta S_{\rm sw}^T({\bf q})$ (see Fig.\ref{geometry}b). The energies have been normalized by the reduced spontaneous magnetization $m_s=M_s(T)/M_s(0)=1.18(1-T/T_c)^{0.36}$\,\, \cite{als-nielsen} and fitted to the predictions of the Holstein-Primakoff theory \cite{HP40},Eq.(\ref{energy}).}}
\label{dispersion}
\end{figure}

A summary of our static results are depicted in Fig.\ref{invint}a, where the inverse ($\omega$-integrated) intensities
of the spin-flip and non-spin-flip channels measured at five different temperatures are plotted against $q^2$. 
The former ones turn out to be independent of temperature and display a clearcut $q^2$-dependence as 
predicted by Eq.(\ref{chisw}) for the spin wave susceptibility $\chi_{sw}$. Note that by using a 
representation of $q_d^2$, which involves the spin wave stiffness $D_{sw}(T)q_d^2=g_L\mu_BM_s(T)$, see
Ref. \onlinecite{Koe83} and Eq.(7) below, one finds an equivalent form, $\chi_{sw}(q \to 0)=g_L\mu_B M_s(T)/D_{sw}(T)q^2$.
In Fig.\ref{invint}a, at the lowest $q^2$ a slight offset is seen, $\chi_{sw}(q=0)=q_d^2/\kappa_g^2$ with $\kappa_g=0.04 \AA^{-1}$.
We ascribe it to the presence of a small gap in the spin wave spectrum, which may be associated with the small cubic
anisotropy of EuS and the finite internal field required to remove the domains.

In contrast to $\chi_{sw}$, the susceptibility of the parallel fluctuations exhibits a strong temperature dependence. 
Figure \ref{invint}a displays the fits of the inverse intensities to Eq.(\ref{chiz}). At large $q^2$, the Ornstein-Zernicke 
behavior is obeyed which by Eq.(5a) defines the inverse ferromagnetic correlation length 
$\kappa_z(T)=\sqrt{18/29}\, \kappa_-(q \gg \kappa_z,T)$, depicted in Fig.\ref{invint}b. Obviously, the
temperature dependence can be well described by the critical law,
\begin{equation}
\kappa_z(T)=\kappa_z(0)(1-T/T_C)^{\nu'}
\end{equation}
with $T_C=16.25(5)$K. The critical exponent $\nu'=0.68(2)$ agrees with the value obtained above $T_C$ of EuS,
$\nu=0.70(2)$\,\, \cite{als-nielsen}, as predicted by the static scaling hypothesis.
For the critical amplitude we obtain $\kappa_z(0)=0.91(5) \AA^{-1}$.
Comparing this value to the amplitude of the paramagnetic correlation length of EuS,  $\kappa_p(0)=0.53 \AA^{-1}$ \,\, \cite{als-nielsen}
we find $\kappa_z=1.7 \kappa_p$, which is bracketed by the mean field value, 
$\sqrt{2} \kappa(0)$ and $2.02 \kappa(0)$\ obtained by considering fluctuations \cite{schinz98a}.

At rather small q, we observe a downward bending of $I_z^{-1}(q)$, which we try to associate with the crossover 
$\chi_z^{-1}(q \ll \kappa_z) \sim q$ following from Eq.(5a). Though the errors are fairly large, we fitted the inverse intensities to 
this prediction and found a=0.20(5), which agrees surprisingly well with a=2/9 predicted by various approaches \cite{schinz98a,mazenko}.
Regarding the fact that the temperature variation of this $I^{-1}(q)$-bending is well 
reproduced, we believe that this constitutes the first signature of the $q^{-1}$-singularity of $\chi_z(q)$ induced by the Goldstone modes.
Interestingly, this variation should also hold for the dipolar critical regime \cite{schinz98a}.

\subsection{Spin wave dispersion}
\label{swdispersion}

The spin wave energies, normalized to the reduced spontaneous magnetization $m_s(T)=M_s(T)/M_s(0)=1.18(1-T/T_C)^{0.36}$\,\,
\cite{als-nielsen,muschke}, are shown in Fig.\ref{dispersion}. For the present orientation of $\bf q$ being perpendicular to ${\bf M}_s$,
spin wave \cite{HP40} and linear response \cite{schinz98b} theories predict for the q-dependence:
\begin{equation}
E(q) = D_0 m_s(T) \tilde{q} ^2 \biggl(1+\biggl(\frac{q_d}{\tilde{q}}\biggr)^2\biggr)^{1/2}.
\label{energy}
\end{equation}
As for the static susceptibility  we admitted the presence of a small gap $\tilde{q} ^2 = q^2 + \kappa_{sw}^2$.
Due to the well-known depolarization of one of the precessing components $\delta{\bf S}_{sw}$, the dipolar wavenumber enters
Eq.(\ref{energy}) via $q_d^2 \equiv g_L\mu_BM_s(0)/D_0$ to cause a crossover from the quadratic dispersion at $q \gg q_d$ 
to the linear law for $q_d \gg q >\kappa_{sw}$. The solid line in Fig.\ref{dispersion} represents the fit of the data to 
Eq.(\ref{energy}) with $D_0= 3.02~meV\AA^2, q_d=0.23(1)~\AA^{-1}$ and $\kappa_{sw}=0.01~\AA^{-1}$.
By comparing this fitted dipolar wavenumber to $q_d=0.22(5)~\AA^{-1}$ as determined from paramagnetic neutron scattering \cite{koetzler86a}
suggests that this quantity is not renormalized by critical fluctuations when passing $T_C$.
Such effect has been conjectured by Fisher and Aharony \cite{fisher73}. However, some indication for
the absence of such renormalizations through critical fluctuations in the dipolar regime has already been realized during a previous
determination of $q_d$\cite{koetzler86}. There, using the {\em mean field} expression for the critical amplitude of the static paramagnetic susceptibility of EuS, 
$C_0=(q_d/\kappa_p(0))^2=0.19$,\,\cite{muschke} $q_d=0.24(2)~\AA^{-1}$ was obtained. 

We also note that $\kappa_{sw}$ is smaller by a factor of four than $\kappa_g$ following from the longitudinal susceptibility and the measured value of $C_0$. We relate this difference to the dipolar interaction, which invalidates the proportionality $\kappa(q) \sim E_{sw}^{-1}(q)$, as can be inferred from the results of Refs.\onlinecite{schinz98c,schinz98a}.

\subsection{Linewidths}
\label{linewidths}
The linewidths evaluated here are defined by the Lorentzian shape which we assumed when analyzing the spectra
by Eq.(\ref{fitfun}). Note that already previous studies on powdered EuS \cite{dietrich} favored this shape over the
Gaussian and the truncated Lorentzian forms. Moreover, also the MMC approach \cite{schinz98c} determined
the damping of the magnetization modes investigated here by assuming an exponential relaxation at long times, which
corresponds to the Lorentzian shape at not too high frequencies.

\begin{figure}
\includegraphics[width=7.5cm]{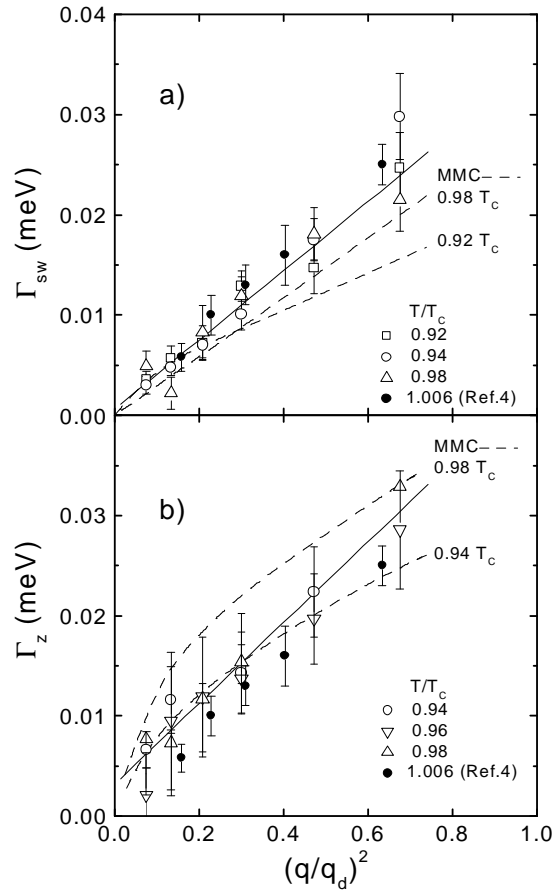}
\caption{\protect {  q-dependence of the Lorentzian linewidths of the a) spin waves and b) longitudinal fluctuations at selected temperatures measured in the dipolar critical regime, see Fig.\ref{geometry}a. For comparison are indicated data taken on powdered EuS at $T=T_C$. }}
\label{widths}
\end{figure}

The results for the widths of the spin wave peaks and of the central peak of the longitudinal fluctuations are displayed
by Fig.\ref{widths}. Within the experimental errors, there is no temperature variation down to the lowest temperatures in region 
DC (see shaded area in Fig.\ref{geometry}a). Note that the errors for $\Gamma_z$ are larger than those for $\Gamma_{sw}$
because of the smaller spectral weight of the longitudinal fluctuations, as it is seen in Fig.\ref{invint}a. As the most
striking result we infer from the presentations against $q^2(<q_d^2)$ in Fig.\ref{widths},  that (i) the relaxation rates of both 
modes obey the simple relations
\begin{subequations}
\label{kincoeff}
\begin{eqnarray}
\Gamma_{sw}(q,T)&=& L_{sw}(\frac{q}{q_d})^2+ \Gamma_{sw}(0,T), 
\\
\Gamma_{z}(q,T)&=& L_{z}(\frac{q}{q_d})^2+\Gamma_{z}(0,T),
\end{eqnarray}
\end{subequations}
being indicated as solid lines in Fig.\ref{widths}, and (ii) that the resulting kinetic coefficients
$L_{sw}=35(2)~\mu eV$ and $L_z=40(3)~\mu eV$ agree within their uncertainties. Moreover, the damping of both
modes remains finite with small values of this background damping, $\Gamma_{sw}(o,T)=0.6(4)~\mu eV$ and $\Gamma_z(q=0)=2.8(4)~\mu eV$, which will also be discussed in the next section.

\section{Discussion}
\label{discussion}
We start from a rather general aspect of the critical phenomena, i.e. the scaling hypothesis extended to dynamical
quantities, like the relaxation rate of the order parameter \cite{halperin}. Then the dipolar interaction in
Heisenberg ferromagnets can fully be taken into account by introducing $q/q_d$ as a second scaling variable
\cite{freylongpaper,schinz98c} in the homogeneous scaling function $g_{\mu}$ for the linewidths of the transverse
modes investigated here, see Fig.\ref{geometry}b:
\begin{equation}
\Gamma_\mu^T(q,T) = q^z \cdot g_\mu\biggl(\frac{q}{\kappa_{\mu}},\frac{q}{q_d}\biggr).
\label{Gamma}
\end{equation}

Let us first approach the problem from the {\em paramagnetic} side to which we designate the index
$\mu=p$. There according to both experiment~\cite{mezei83} and MMC theory~\cite{frey88} the
crossover from exchange dominated dynamics, being characterized by the exponent $z=(D+2)/2=5/2$,
to the dipolar critical dynamics with $z=2$ occurs deep in the DC region, i.e. for $q, \kappa(T)\ll q_d$.
This is the reason, why dipolar effects were not realized in the early investigations~\cite{collins,minciewicz,dietrich,bohn}.
Once the dipolar dynamics, i.e. $z=2$, has taken over very close to $T_C$, where $\kappa_p(T)  \ll q_d$,
the paramagnetic scaling function assumes the constant value (see. Eq.(7) of Ref.\onlinecite{koetzler88}):
\begin{equation}
g_p(\infty,0) = L_d \,q_d^{-2}.
\end{equation}
For EuS, the (dipolar) kinetic coefficient $L_d$ has been determined from the analysis of the relaxation rates 
of the transverse fluctuations in the limit $q=0$ above \cite{KKW76} and below $T_C$ \cite{goekoe98}:
\begin{equation}
\Gamma^T_p(q \to 0,T)=\frac{L_d}{\chi_p^T(q\to 0)}
\end{equation}
The value $\hbar L_d=38(4)~\mu eV$ was found to agree fairly well with the MMC estimate \cite{frey88,freylongpaper,koetzler88}
$\hbar L_d=5.1g_L\mu_B\sqrt{k_BT_C/\mu_0}\,q_d^{3/2}/4\pi^2 \simeq 30~\mu eV$.

{\em Below $T_C$}, as a matter of fact, we recover here the same 
$q^2$-dependence of both relaxation rates $\Gamma_z$ and $\Gamma_{sw}$, Eq.(\ref{kincoeff}), and in particular, within the experimental errors,
their coefficients agree with $L_d$,  $L_{sw} \approx L_z \approx L_d$.
These are the central results of our work. In terms of the dynamical scaling hypothesis they imply that by passing 
the Curie temperature from above, the dipolar dynamic universality class for the transverse (critical) fluctuations $\delta{\bf S}^T$ 
characterized by $z=2$ is not changed. 
This basic feature is nicely confirmed by (rather old) $\Gamma_p^T({\bf q})$ data measured slightly above $T_C$ of powdered EuS, which we have added to Fig.\ref{widths}a. Within the given error bars, they agree with respect to magnitude and $q^2$-variation with the present $\Gamma_{sw}^T$ data taken below $T_C$. We also observe a slight systematic enhancement of $\Gamma_z^T(q,T<T_C)$ over $\Gamma_p^T(q,T_C)$, which seems to be associated with the (small) background $\Gamma_{z,bg}$ to be discussed below.

To date, the fundamental phenomenon of dynamical scaling in Heisenberg ferromagnets has been established
for the exchange ('true')critical regime (EC in Fig.\ref{geometry}a), i.e. for $q>\kappa(T)>q_d$, where $\Gamma^T$ becomes independent of temperature, i.e.
$\Gamma^T \sim (q/q_d)^{5/2}$ is maintained on both sides of $T_C$ \cite{collins,minciewicz,dietrich,bohn,koetzler88}. Our experiment provides evidence for the 'true' (temperature independent) dipolar critical behavior extending to even for smaller q-values,
$q<\kappa_z(T)$, see shaded region in Fig.\ref{geometry}a. A temperature variation might set in when the dipolar hydrodynamic (DH) is reached.  This extended {\em dynamic} critical behavior
with $z=2$ below $T_C$ is also very much different from the situation above $T_C$, where $z=2$ is attained
for extremely small $q \ll q_d$\,\,\cite{mezei83,frey88}, and only a small part of the static DC region is 
occupied by {\em dynamic} dipolar criticality. Both features are consistent with the general fact that the dynamic critical
behavior reflects more details of the system, in particular conservation laws \cite{halperin}, than the static properties, 
like the susceptibility tensor $\tensor{\chi}(\bf q)$, which define the critical regimes displayed in Fig.\ref{geometry}a.

Having established the dynamic class, $z=2$, in a large part of DC below $T_C$, we are now able to discuss the
further consequences of the q-variation of the damping, Eq.(\ref{kincoeff}). The kinetic coefficients $L_z$ and $L_{sw}$
turned out to be identical for both the spin wave and the longitudinal modes. By looking at Fig.\ref{geometry}b, this result 
emerges from the continuity of the critical behavior of the transverse modes $\delta{\bf S}^{T_1}({\bf q})$ and
$\delta{\bf S}^{T_2}({\bf q})$.  
Note that above $T_C$, the designations 'transverse' and 'longitudinal' define orientations of $\delta{\bf S}$ with respect to ${\bf q}$ being indicated by a superscript in $\delta{\bf S}^\alpha$. Below $T_C$, due to the symmetry breaking through $M_s{\bf e}_z$, this 'conventional' definition of the mode polarization $\delta{\bf S}({\bf q})$ is possible only for certain ${\bf q}$-directions, like the one chosen in the present work.
The 'transverse' modes drive the ferromagnetic transition on the paramagnetic side, where the {\bf q}-vector
clearly determines the symmetry. By passing $T_C$ from above, Fig.\ref{geometry}b suggests the continuous transformations,
$\delta{\bf S}^{T_1} \to \delta{\bf S}^{T}_{sw}$ and $\delta{\bf S}^{T_2} \to \delta{\bf S}^{T}_{z}$.  Note that this is only true 
for our experimental configuration, where the order parameter ${\bf M}_s$ is oriented perpendicular to ${\bf q}$. The fact
that the kinetic coefficients below $T_C , L_{sw}$ and $L_z$ are equal and, moreover, agree with $L_p=38(2) \mu eV$ of
$\delta{\bf S}^T({\bf q})$ above $T_C$\, \cite{bohn,koetzler88} (see also Fig.\ref{widths}, can be immediately be related to the modes with
transverse polarization $\delta{\bf S} \perp {\bf q}$, which are the only ones to become critical in real Heisenberg ferromagnets,
i.e. with dipolar interaction. On the paramagnetic side, these critical modes display relaxational dynamics, but below $T_C$ their dynamical shape depends on the direction of their propagation vector with respect to the order parameter ${\bf M}_s$. If these transverse critical modes propagate, for example, along ${\bf M}_s$ they are predicted to exhibit spin wave dynamics in DC \cite{schinz98b,schinz98c} with pure Goldstone-like susceptibilities $\chi_{sw}^T({\bf q} \parallel {\bf M}_s) =(q_d/q)^2$\,\, \cite{schinz98a}. The remaining longitudinal mode, $\delta{\bf S}_z^L \parallel {\bf q}$, should be strongly damped and suppressed in intensity. These modes can be studied in a configuration, where the polarizing field $\bf B$ is oriented parallel to the scattering plane.

The dipolar symmetry with respect to $\bf q$ is also reflected by the fact that in DC the direction of $\bf q$ is parallel to the largest
eigenvector of the susceptibility tensor $\tensor{\chi}({\bf q})$ \cite{koetzler88}, ${\bf v}_3$, while the second largest, ${\bf v}_2$, is
parallel to ${\bf M}_s$. This symmetry of $\tensor{\chi}({\bf q})$ changes if the DC-regime is left. Then the directions of these
two eigenvectors are just interchanged for our configuration ${\bf q} \perp {\bf M}_s$, while for a general orientation between ${\bf q}$
and ${\bf M}_s$ a gradual rotation, ${\bf v}_3 \to {\bf v}_2$ and ${\bf v}_2 \to -{\bf v}_3$, takes place. Hence, outside of DC,
the 'leading' static symmetry of the ferromagnet is defined by the order parameter ${\bf M}_s$ and, moreover, the dynamical exponent
attains its isotropic value, $z=5/2$. This implies that there both spin wave modes, $\delta{\bf S}^T_{sw} \perp {\bf q}$ and 
$\delta{\bf S}^L_{sw} \parallel {\bf q}$, are critical. Recently, their frequencies have been investigated in some detail by polarized neutrons \cite{boni95}.
For the damping similar data are still lacking. Early work employing unpolarized neutrons \cite{dietrich}  away from $T_C$ of EuO provided relaxation rates, which where consistent with
$\Gamma_{sw}(q,T) \sim \kappa^{-3/2}_zq^4$. This agrees with the scaling hypothesis, Eq.(\ref{Gamma}),
provided $g_{sw}=(q/\kappa_z(T))^{3/2}\tilde{g}_{sw}(0,\infty)$. Unlike our observation in the DC region, in the exchange critical 
regime the dominance of the thermally excited spin waves gives rise to the strong increase of their linewidths with temperature.
The leading $q^4$-dependence has been predicted by Vaks et al.\cite{VLP68} and results from spin wave - spin wave scattering.

Another interesting point is the fact, that both linewidths do not change with temperature down to the lowest values studied here, 
$T/T_C=0.92$. 
This is somewhat surprising with regard to the existing results of the MMC-calculations \cite{schinz98c,schinz94}, which using some interpolation have also been indicated in Fig.\ref{widths}. We notice that they predict a slight temperature variation, i.e. a narrowing with decreasing $T$ which is not observed. Also the $q^2$-variations of our linewidths are not reproduced and the absolute MMC values for $\Gamma_{sw}$ and $\Gamma_z$ are smaller and larger, respectively. This seems to indicate that the assumptions of the MMC approach, like the Lorentzian  approximation for the modes with large $q$ and some cutoff of the dynamics, may not be valid. By a more phenomenological point of view, we rather conjecture that the non-critical longitudinal fluctuations, $\delta{\bf S}^L_z \parallel {\bf q}$, play a much more important role than believed to date. Their damping is expected to be $\Gamma_z^L({\bf q})=L_d/\chi_z^L=L_d$, i.e. temperature independent and rather large and may provide an efficient relaxation channel for the critical modes. The importance of $\Gamma_z^L$ has been realized recently in the relaxation rate of the homogeneous magnetization $\Gamma_z^T$ below $T_C$\, \cite{goekoe98}.

A final comment on the 'background' damping of the longitudinal fluctuations, $\Gamma_z(q=0,T \le T_C)=2.8 \mu eV$ may be appropriate.
We conjecture here that - similar as above $T_C$ \,\cite{koetzler88} - damping effects by the longitudinal polarization of the $\delta{\bf S}^L_z$ mode are picked up within the resolution of our experiment. Due to the dipolar demagnetization, these modes are
uncritical ( dynamic exponent z=0 \,\cite{freylongpaper}). This conjecture is based on the fact that our background value is rather close to $\Gamma_{z,bg}=L_{bg}/\chi_z^L(q=0,T \le T_C)=1.8(2) \mu eV$, where we have inserted
(i) the background kinetic coefficient $L_{bg}=1.8(2) \mu eV$ determined from the critical behavior of $\Gamma^T(q=0,T>T_C)$
\cite{koetzler88}, and (ii) the susceptibility of the longitudinal spin wave modes, $\chi_{sw}^L(q \to 0,T<T_C)=1$, which at $T=T_C$
transform into the longitudinal paramagnetic modes $\chi^L(q \to 0)$.  It may be interesting to note that this background seems to be absent or at least significantly be reduced in the paramagnetic relaxation rate $\Gamma_p^T({\bf q},T)$ of Ref.\onlinecite{bohn}, depicted in Fig.\ref{widths}. This indicates that the appearance of the order parameter increases $\Gamma_z^T(0,T)$ while its effect on the background rate of the spin wave linewidth, $\Gamma_{sw}^T(0,T)=0.5(4)~\mu eV$ turns out to be small.

\section{Summary and Conclusions}
\label{summary}
We have conducted a study of the dynamics of spin fluctuations in the ferromagnetic state of EuS close to $T_C$, where dipolar effects are expected to play a dominant role. For intensity reasons we concentrated on the spin wave and longitudinal fluctuations, $\delta{\bf S}^T_{sw}$ and $\delta{\bf S}^T_{z}$, both transverse with respect to the momentum transfer $\bf q$, which was chosen perpendicular to $\bf M$. This configuration allowed to extract from the spin wave frequencies the characteristic dipolar wavenumber $q_d$ and from the static susceptibility, $\chi_z({\bf q},T)$ the correlation length $\xi_z=\kappa_z^{-1}$ of the longitudinal fluctuations. The latter exhibits deviations from the Ornstein-Zernicke law at $q \ll \kappa_z$, which appear to be related to the predicted thermal renormalization of $\chi_z$ by the spin waves \cite{mazenko,schinz98c}. This feature, though subject of rather intense research in the past years (see e.g. References \onlinecite{mitchell,boni91a,coldea}), has not been identified before.
The reason for the observation of the mass renormalization of $\chi_z({\bf q},T)$ on EuS can be attributed to the rather large critical amplitude $\kappa_z(0)=0.91 \AA^{-1}$, which allows to explore the regime $q/\kappa_z \ll 1$. This is in contrast to the situation in the itinerant ferromagnet Ni\,\cite{boni91a}, where $\kappa_z(0)$ is almost one order of magnitude smaller.

As the central result of our study, we regard the linewidths of both modes. They display the same q-variation $\Gamma_{\mu} = L_d (q/q_d)^2$, which we explained by using the dynamical scaling hypothesis and existing data for $\Gamma^T({\bf q},T \ge T_C)$ of EuS. Apart from a small finite background for $\Gamma_z(q=0)$, which was already observed above $T_C$ in a previous work, the absolute values of both linewidths prove to be identical. Moreover, the relevant kinetic  coefficient $L_d$ agrees with the value obtained from the linewidths of the transverse fluctuations $\delta {\bf S}^T_p({\bf q})$ measured in the dipolar critical regime slightly above $T_C$. This quantitative feature suggests the transformations
$\delta{\bf S}^{T_1}_p \to \delta{\bf S}^{T}_{sw}$ and $\delta{\bf S}^{T_2}_p \to \delta{\bf S}^{T}_{z}$ when passing $T_C$ from the para- to the ferromagnetic side without any change of the dynamics and thus obeying dynamical scaling - as conjectured in Ref.\onlinecite{sacavem}. As a surprising feature we note that the 'true', i.e. temperature independent dipolar critical behavior extends so far into the static DC region explored here, see Fig.\ref{geometry}a. This contrasts to the extremely narrow regime just above $T_C$, in which the dipolar anisotropy changes the dynamic universality class of the transverse modes from $z=5/2$ to $z=2$. The observed q- and temperature variations of our linewidths are not fully consistent with predictions by the MMC-theory \cite{schinz98c}, which is a bit unexpected regarding the success of this approach on the paramagnetic side.\cite{freylongpaper,schinz98b}

Finally we should note that for intensity reasons we were not yet able to measure the crossover of $\Gamma_z^T$  to the hydrodynamic $(q < \kappa_z/9)$ and to the exchange critical $(q> q_d)$ regime, for which some more theoretical work has been published.
\cite{resibois-piette,maleev,mazenko,VLP68,freylongpaper} An even greater experimental challenge is the low intensity of the longitudinal polarizations $\delta{\bf S}_{sw}^L$\,\cite{sacavem} and $\delta{\bf S}_z^L$. Their non-critical dynamics may be responsible for the absence of any temperature effects on the both linewidths investigated here. Other unsettled problems are the mechanisms responsible for the small and non-critical background effects occuring in the static susceptibilities, spin wave frequencies and relaxation rates. The values quoted here for $\kappa_g, \kappa_{sw}, \Gamma_z(0,T)$ and $\Gamma_{sw}^T(0,T)$ may help to answer the question whether there exists a common origin, like anisotropy or finite internal magnetic field.


\begin{thebibliography}{80}
\bibitem{background} The energy independent background of 1
                 count/8.3 min seems to be unreasonably small. The reason is that
                 the fitted spectral weight functions were approximated by
                 Lorenzians that usually enforce a too low background when fitting
                 data as discussed in Ref. \onlinecite{boni87}.
\end{thebibliography}
\end{document}